%% file: main.tex
\documentclass[journal]{IEEEtran}
\usepackage[latin1]{inputenc}
\usepackage{times,amsmath}
\usepackage{amssymb}
\usepackage{steinmetz}
\usepackage{pstool}
\usepackage{subfigure}
\usepackage{multirow}
\usepackage{enumerate}
\usepackage{graphicx}
\usepackage{multirow}
\usepackage{subfig}
\usepackage{booktabs}
\usepackage{mwe}
\usepackage{bigints}
\usepackage{MnSymbol}
\usepackage{stfloats} 
\usepackage[table]{xcolor}
\usepackage[square, comma, sort&compress, numbers]{natbib}
\usepackage{nohyperref}
\usepackage{algorithm,algorithmic}
\usepackage{epstopdf}
\usepackage{mathtools, cuted}

\graphicspath{ {Figures/} }

\begin{document}

%\title{ Radio-PPG: Digital twin of PPG acquired in a non-contact manner using RF Data Collected of the Chest}
%\title{ Radio-PPG: Non-Contact PPG monitoring using RF Data Collected of the Chest}
%\title{Non-Contact Acquisition of PPG Signal using Radio Data Collected via Software-defined Radio}
\title{Non-Contact Acquisition of PPG Signal using Chest Movement-Modulated Radio Signals}

\author{
\IEEEauthorblockN{
Israel Jesus Santos Filho\IEEEauthorrefmark{1}, Muhammad Mahboob Ur Rahman\IEEEauthorrefmark{1}, Taous-Meriem Laleg-Kirati\IEEEauthorrefmark{2}, Tareq Al-Naffouri\IEEEauthorrefmark{1} }

\IEEEauthorblockA{\IEEEauthorrefmark{1}Computer, Electrical and Mathematical Sciences and Engineering Division (CEMSE), \\ 
King Abdullah University of Science and Technology, Thuwal 23955, Saudi Arabia.\\ 
\IEEEauthorrefmark{2}The National Institute for Research in Digital Science and Technology, Paris-Saclay, France.\\
\IEEEauthorrefmark{1}\{israel.filho,muhammad.rahman,tareq.alnaffouri\}@kaust.edu.sa, \IEEEauthorrefmark{2}Taous-Meriem.Laleg@inria.fr }
}

%Computer, Electrical and Mathematical Sciences and Engineering Division (CEMSE), King Abdullah University of Science and Technology, Thuwal 23955, Saudi Arabia.

\maketitle

\input{abstract}

\input{sec1}

\input{sec2}
\input{sec3}

\input{sec4}
\input{conclusion}

%\input{Appendix1}

%\appendices

%\input{appendix}

%\section*{Acknowledgements}

\footnotesize{
\bibliographystyle{IEEEtran}
\bibliography{references}
}

\vfill\break

\end{document}

%% file: abstract.tex
\begin{abstract} 

We present for the first time a novel method that utilizes the chest movement-modulated radio signals for non-contact acquisition of the photoplethysmography (PPG) signal. Under the proposed method, a software-defined radio (SDR) exposes the chest of a subject sitting nearby to an orthogonal frequency division multiplexing signal with 64 sub-carriers at a center frequency 5.24 GHz, while another SDR in the close vicinity collects the modulated radio signal reflected off the chest. This way, we construct a custom dataset by collecting 160 minutes of labeled data (both raw radio data as well as the reference PPG signal) from 16 healthy young subjects. With this, we first utilize principal component analysis for dimensionality reduction of the radio data. Next, we denoise the radio signal and reference PPG signal using wavelet technique, followed by segmentation and Z-score normalization. We then synchronize the radio and PPG segments using cross-correlation method. Finally, we proceed to the waveform translation (regression) task, whereby we first convert the radio and PPG segments into frequency domain using discrete cosine transform (DCT), and then learn the non-linear regression between them. Eventually, we reconstruct the synthetic PPG signal by taking inverse DCT of the output of regression block, with a mean absolute error of 8.1294. The synthetic PPG waveform has a great clinical significance as it could be used for non-contact performance assessment of cardiovascular and respiratory systems of patients suffering from infectious diseases, e.g., covid19.

%Then we compared the capabilities of each of three signals in the task of estimation of the vital signs, the oxygen saturation and heart rate. Our results showed similar capabilities by using the raw ppg waveform and the radio-ppg generated by our algorithm, as well the raw-sdr waveform for estimating the vitals., and an MAE of 5.645 for the oxygen saturation and 8.715 for beats/min estimation problem through the radio-ppg. We are the first work to explore the capabilities of digital twins representation of biological signals by leveraging RF sensing with SDR providing not only a methodology to collect and process the signals, whilst their application.

\end{abstract}


%% file: sec1.tex
\section{Introduction}
\label{sec:intro}

Photoplethysmography (PPG) is a simple, cost-effective and non-invasive method that utilizes optical principles to measure rhythmic changes in the blood volume at a peripheral location. PPG signal has recently gained widespread attraction among the researchers and clinicians alike, as a viable biomarker for monitoring of vital signs, and for performance assessment of cardiovascular and respiratory systems \cite{almarshad2022diagnostic}. PPG devices come in various forms, e.g., wearables, smart watches, smart bands, smart earphones, smartphones, consumer electronics such as cameras, etc. \cite{ppg_pot_2018}. PPG, being a non-invasive method, allows seamless and continuous health monitoring in various diverse settings, from in-hospital patient monitoring to well-being of athletes outdoors, to sleep quality analysis indoors. The ease-of-use and affordability contribute to the scalability of PPG-based health monitoring, potentially helping it reach a broader population and facilitating patient-centric healthcare \cite{almarshad2022diagnostic}. PPG waveform contains a wealth of information that helps in estimation of a number of biomarkers, e.g., body vitals \cite{mehmood2023your}, blood pressure  \cite{tahir2024cuff, li2021central} etc., and diagnosis of a number of diseases, e.g., vascular aging \cite{saran2023low}, atrial fibrillation and more. 

%\cite{ahmedtahirpaper, li2021central, kurylyak2013neural}, vascular aging \cite{saranpaper, dall2020prediction, charlton2022assessing}, and more.

More recently, researchers have started developing methods for acquiring the PPG signals from a distance---the so-called remote or non-contact PPG acquisition methods \cite{remote_ppg_1_2020}. Such methods mostly utilize a standard RGB camera to record a small video of the face from a distance. This video is known to capture the subtle-but-periodic changes in the skin color of the face due to cardiac activity, which helps extract the PPG signal \cite{kumar2015distanceppg}. Very recently, researchers have successfully demonstrated cardiac pulse detection using another sensing modality, i.e., a terahertz (THz) transceiver, by measuring the reflectance of THz waves from upper dermis tissue layer of the skin \cite{rong2022new}.

Another set of closely related works include methods that aim to solve a more broad range of health sensing problems in non-contact manner, using many different radio frequency (RF) sensing modalities. Such methods gained popularity in the post-covid19 era, starting with body vitals estimation of covid19 patients from a distance \cite{taylor2020review}. But more lately, such methods are being developed for a multitude of applications e.g., fall detection, sleep stage analysis, gait analysis, etc. 

RF-based health sensing methods could be mainly categorized into three kinds: 1) Radar-based methods, 2) Software-defined radio (SDR)-based methods, 3) WiFi-based methods. The radar-based sensing methods deploy various kinds of radars (e.g., ultra-wideband pulse radar, frequency modulated continuous-wave radar, etc.) that assess cardio-respiratory performance of a person in a contactless manner by means of the classical range and Doppler analysis \cite{ahmed2023machine}. The SDR-based sensing methods, on the other hand, utilize the fluctuations in the amplitude, frequency and phase of the microwave-band signals reflected off the human body to measure vitals and respiratory abnormalities \cite{pervez2023hand,buttar2023non}. Finally, Wi-Fi-based sensing methods exploit the extensive existing infrastructure of WiFi routers indoors, and utilize machine and deep learning algorithms on the signals reflected by the human subjects, in order to do various tasks, e.g., activity recognition, sleep analysis, fall detection, etc., to help realize a smart home \cite{ge2022contactless}.

Non-contact methods could improve the comfort-level of the patients, decrease the infection risks, and offer continuous, seamless and in-situ health monitoring. Further, non-contact methods could aid in early disease diagnosis, could enable personalized and proactive healthcare solutions, could reduce the burden on existing healthcare systems, and thus could help realize the vision of smart homes and smart cities of future \cite{taylor2020review}. Nevertheless, these methods have their pros and cons also. For example, these methods need to deal with a range of issues such as motion artifacts, diverse environmental conditions, complex body movements, \cite{ahmed2023machine} etc. Last but not the least, camera-based remote PPG methods are considered to be a privacy-breach, and this slows down the research in this direction \cite{remote_ppg_1_2020}.

%This work lies in this kind of scheme to sensing and we are not focused on the monitoring of human body, but in the capabilities of wave translation from the information collected by the reflected waves from the body. 

%This work is first study that uses RF-waves that produces good quality PPG in a non invasive way and rely on our signal processing pipeline to justify the use of machine learning models in more controlled way. Our application of deep learning for waveform conversion in PPG is not only limited to noise reduction but can also involve transforming signals between different sensor modalities (through the of different channels of our OFDM system, but the system can be extended to incorporate another sources of information like radar, lidar and other sensors measurements in order to be more robust) or adapting to variations in hardware and environmental conditions. To conclude, this work presents a combination of signal processing techniques and machine learning resulting in a novel way to collect and apply rf-sensing to synthesize biomarkers and do vital signs estimation.

{\bf Contributions.} 
This work belongs to the umbrella of RF sensing methods, and capitolizes on the chest movement-modulated radio signals for non-contact acquisition of the PPG signal. The two key contributions of this work are as follows:
\begin{itemize}
    \item We construct a custom labeled dataset---first of its kind---by simultaneously collecting 2.5 hours worth of raw radio data and reference PPG data from 16 subjects.
    \item We present a novel method that consists of a purpose-built data pre-processing + deep learning pipeline which efficiently learns the PPG signal representation from the radio signal, in a contactless manner. 
\end{itemize}

{\it To the best of our knowledge, this is the first work that does non-contact PPG monitoring using the radio signals.}

{\bf Outline.} 
Section II provides a compact discussion of the data collection process. Section III describes the proposed data pre-processing and regression pipeline. Section IV discusses selected results. Section V concludes the paper.

%% file: sec3.tex
\section{The Experimental Setup \& Data Acquisition}
\label{sec:sys-model}

Given the fact that there currently exists no dataset that simultaneously records radio data and PPG data of healthy human subjects, we construct one such dataset with the aim to learn the mapping between the radio signal and the PPG signal. Below, we discuss the experimental setup followed by the necessary details about the dataset that we have constructed. 

\subsection{The Experimental Setup}

\begin{figure}[ht]
\begin{center}
	\includegraphics[width=\linewidth]{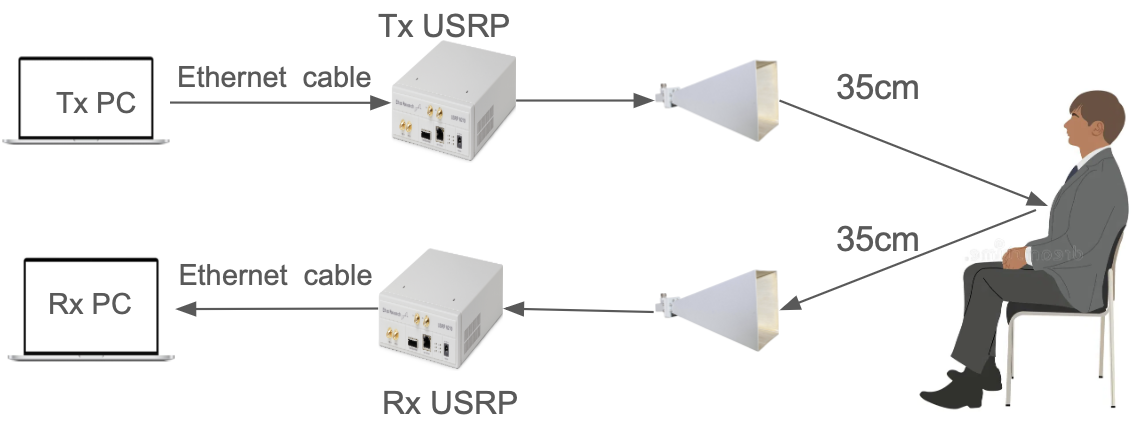} 

\caption{\footnotesize The proposed method for non-contact PPG monitoring: The experimental setup consists of an SDR pair whereby the transmit SDR exposes the chest of a subject with an OFDM signal, while the receive SDR collects the reflected signal and feeds it to a purpose-built data pre-processing + deep learning pipeline that ultimately synthesizes the PPG signal.}
\label{fig:sysmodel}
\end{center}
\end{figure}

The proposed method for non-contact acquisition of the PPG signal utilizes two software-defined radios (SDR), each connected with a workstation by means of an Ethernet port (see Fig. \ref{fig:sysmodel}). Specifically, we utilize a pair of Universal Software Radio Peripheral (USRP) N210 SDRs, each connected with a directional horn antenna, which forms a single-input, single-output (SISO) link. We utilize MATLAB R2021a to program both the transmit and receive USRP SDRs as follows. The transmit SDR sends an orthogonal frequency division multiplexing (OFDM) signal consisting of 64 sub-carriers, with quadrature phase shift keying (QPSK) modulation on each sub-carrier. Eventually, the transmit SDR appends a cyclic prefix (CP) of duration 16 samples to each OFDM symbol. The receive SDR removes the CP from the received signal, and logs the complex-valued time-domain data which is later pre-processed offline and fed to the proposed method that ultimately synthesizes the PPG signal. For the experiment, we set the gain of the transmit horn antenna to 40 dB. Further, we utilize a center frequency of 5.24 GHz, a baseband sampling rate of 20K samples/sec, and a bandwidth of 20 KHz.

%From the received signal, the receiver derives the CFR that was eventually utilised to detect the dehydrated subject using the ML classifiers.
% Fig. \ref{fig:sysmodel} shows the system model (flowchart) for our proposed SDR-based, ML-empowered system for detection of breathing abnormalities. 

\subsection{Data Acquisition for the Radio-PPG dataset}
%Extensive studies are run to assess the platform's performance in identifying and categorising the dehydrated subject after the transmitter and receiver have been designed.

Having described the key details of the experimental setup, we now discuss the pertinent details of the data acquisition process. Fig. \ref{fig:sysmodel} illustrates the working of the proposed method for non-contact synthesis of PPG waveform. As can be seen in Fig. \ref{fig:sysmodel}, the subject sits nearby the SDR pair (roughly 35 cm away), the transmit SDR strikes the OFDM signal onto the chest of the subject, while the receive SDR collects the signal reflected off the chest of the subject.\footnote{This study was approved by the ethical institutional review board (EIRB) of KAUST, Saudi Arabia.} For the purpose of ground truth acquisition, we utilize MAX86150 module that allows us to acquire the reference PPG signal at a sampling rate of 2.5 KHz when a subject places his/her finger on the on-board PPG sensor of the MAX86150 module.

We construct the custom {\it Radio-PPG-16 dataset} by collecting data from 16 volunteers (10 male, 6 female, aged 25 to 32 years). For each subject, we do two measurement sessions, whereby in each session we simultaneously collect RF data and PPG data for a duration of 5 minutes\footnote{This simultaneous collection of radio data and PPG data helps us synchronize the synthetic PPG signal with the reference PPG signal later on.}. During each experiment session, we make sure that the subject sits still in order to avoid motion-induced artefacts in the data being gathered. In total, we collect 10 minutes of data per subject, and 160 minutes of labeled data from the 16 subjects.

%% file: sec4.tex
\section{Non-Contact PPG Synthesis}
The proposed method consists of two distinct phases, i.e., data pre-processing and waveform translation (regression). 

\subsection{Data Pre-processing Phase}  

Data pre-processing phase prepares the data for the deep learning models that translate the radio signal to PPG signal. This phase consists of the following steps: 1) pre-processing of the reference PPG data, 2) pre-processing of the radio data, and 3) synchronization of the reference PPG and radio data.  

\subsubsection{Pre-processing of reference PPG data}

The reference PPG signal (acquired through MAX86150 module) suffers from a number of distortions, e.g., respiration-induced baseline drift, muscle interaction-induced artefacts, artefacts due to variations in ambient light conditions, motion artefacts, etc. To this end, we manually identify and remove the artefacts (of various origins) throughout the dataset. Further, we utilize wavelet transform---the {\it db2} wavelet family---to estimate and remove the baseline drift. Subsequently, we utilize a 12-th order Butterworth low-pass filter with cut-off frequency 3.4 Hz, in order to remove out-of-band noise and retain only the frequency components representing blood volume changes. We then normalize the conditioned PPG data using Z-score normalization method. Finally, we segment the data in 2.2 seconds long non-overlapping segments.

\subsubsection{Pre-processing of Radio data}
Pre-processing of the radio data consists of the following main steps:
\begin{itemize}
    \item First of all, capitalizing on the frequency diversity rule and inline with prevailing OFDM channel estimation practices, we proceed to work with radio data from 16 equi-spaced sub-carriers, out of 64 sub-carriers.\footnote{The two frequency responses of the human body when exposed to two frequencies that are quite close to each other, are also very similar.} 
    \item Next, we proceed to fuse the length-$N$ radio data of 16 sub-carriers ($X \in \mathbb{C}^{16 \times N}$) in a linear fashion in order to do dimensionality reduction, through the principal component analysis (PCA). Since the PCA is not defined for complex signals, we surrogate this problem by computing the PCA for the real part and imaginary part of the data separately, to get first principal component $\hat{X}_R\in \mathbb{R}^N$ and first principal component $\hat{X}_I\in \mathbb{R}^N$, respectively. We then append the imaginary part $\hat{X}_I$ to the real part $\hat{X}_R$ to get $\hat{X}\in \mathbb{R}^{2N}$. Eventually, we apply the modulus operator to get $|\hat{X}|\in \mathbb{R}_+^{2N}$.
    \item We then apply discrete wavelet transform (DWT) on $|\hat{X}|$ to obtain a 10-level wavelet decomposition. This allows us to remove artefacts, out-of-band noise, and frequency components that do not represent blood volume changes, by zeroing out the corresponding detailed and approximation wavelet coefficients during the wavelet reconstruction phase.
\end{itemize}
We then normalize the conditioned radio data using Z-score normalization method. Finally, we segment the data in 2.2 seconds long non-overlapping segments. 

%define: $X = X_{real} + jX_{imag}$, where j stands for the complex unity. Using this formulation, we can compute the PCA for the real and imaginary components of X, $X_{real}, X_{imag}$, and project the whole set into the first component, resulting in $\hat{X} = \hat{X}_{real} + j\hat{X}_{imag}$. 

%It is necessary to be sure that the sdr signal contains the frequency components associated with the required biological markers. In order to analyse it we applied the discrete wavelet decomposition to decompose the signal into high and low frequency components . The target frequency lies in the range of 2 to 3.5Hz, to achieve this, and based on our USRP hardware with sample frequency of 2.5KHz, we applied a wavelet tree with depth of 10, meaning that in the last 3 levels we are able to analyse from ~2.4Hz to ~9.4Hz. In the wavelet decomposition, the detail coefficients are the low frequency signals and the approximation coefficients are the high frequency signals in each level of the tree, like fig \ref{fig:alig_sig} and \ref{fig:non_alig_sig}.

\subsubsection{Peak synchronization}
\begin{figure}[ht]
\begin{center}
	\includegraphics[width=9.5cm,height=5.7cm]{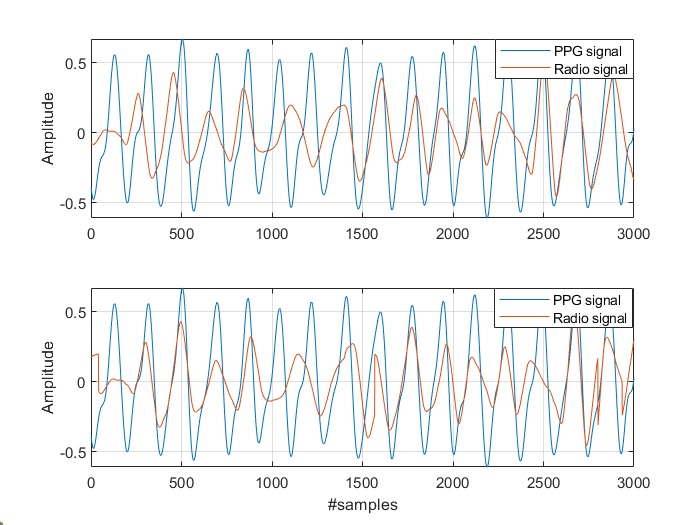} 
\caption{\footnotesize Top fig. shows a pair of radio and PPG segments after DWT stage but before alignment, while the bottom fig. shows the situation after alignment.}
\label{fig:peak_sync}
\end{center}
\end{figure}

Synchronization of the radio and PPG signals is an essential data pre-processing step that ensures one-on-one mapping between the radio signal and PPG signal, at the sub-cardiac cycle resolution. We synchronize the radio and PPG segments by maximizing the alignment between them in a peak-to-peak fashion (i.e., by minimizing the phase offset/mismatch between the two segments). We achieve this by means of the cross-correlation approach, i.e., we sweep over a wide range of time shift values in the radio segment such that its inner product with the corresponding reference PPG segment is maximized. The inner product, being a viable normalized similarity measure, is maximized when both segments are super-imposed, resulting in a peak-to-peak alignment between the two segments. Fig. \ref{fig:peak_sync} shows that the miss-aligned peaks of the radio and PPG segments (see top fig.) become aligned due to cross-correlation based peak synchronization method (see bottom fig.).

\begin{figure*}[ht]
\begin{center}
	\includegraphics[width=18cm,height=4.5cm]{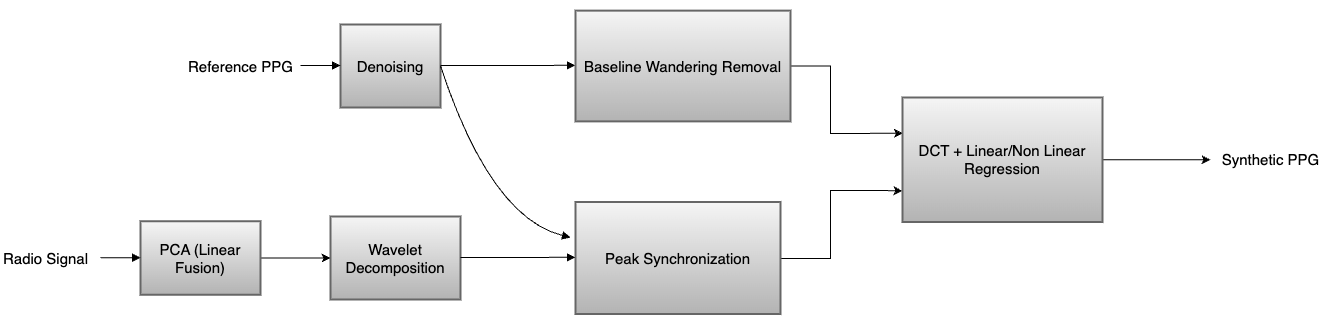} 
\caption{\footnotesize Complete data pre-processing + regression pipeline for PPG waveform synthesis from the radio signal.}
\label{fig:complete_pipalig_sig}
\end{center}
\end{figure*}

%updated_processing_flow.png

\subsection{Waveform Translation Phase}

We now discuss the regression mechanism that helps us translate the conditioned and pre-processed radio signal into a PPG signal, during the test phase. In this work, we do the regression in frequency domain. That is, during the training phase, we begin by computing the type-II discrete cosine transform (DCT) of both the reference PPG and radio segments. With this, we implement the following two methods to learn the regression between the DCT coefficients of the ground truth PPG and DCT coefficients of radio signal:
1) linear (ridge) regression, 2) non-linear regression by means of a multi-layer perceptron (MLP) with 5 layers. 
This way, during the test phase, the proposed method returns us the DCT coefficients of the PPG signal, in response to an input radio signal. Eventually, we obtain the synthetic PPG signal by taking the inverse DCT during the test phase. 

Fig. \ref{fig:complete_pipalig_sig} shows the the block diagram that illustrates the complete data pre-processing + regression pipeline of the proposed method, as discussed above.

\section{Results}

We begin by summarizing the key hyperparameters of the proposed method, followed a discussion of the selected results. 
 
\begin{figure}[ht]
\begin{center}
	\includegraphics[width=7cm, height=6cm]{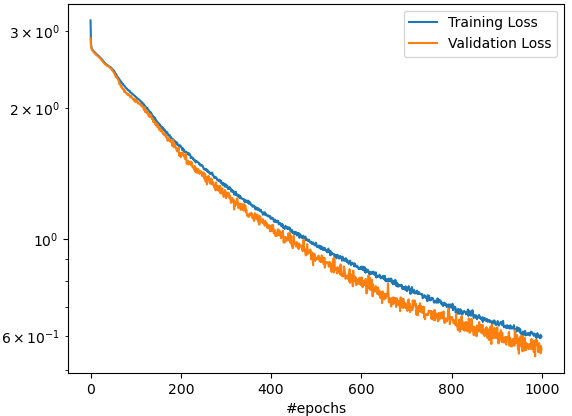} 
\caption{\footnotesize Training and validation losses of the proposed MLP regressor undergo a steady decrease with the increase in number of epochs.}
\label{fig:loss_epoch_curve}
\end{center}
\end{figure}

We utilize 80\% of data for training, and remaining 20\% data for testing purpose. Next, we compute the DCT and inverse DCT at the segment level (each radio and PPG segment consists of 400 samples). For the MLP regressor, we use leaky-relu activation function in all layers except the last layer. For backpropagation purpose, we use mean absolute error (MAE) as the loss function. Further, in order to avoid overfitting, we use L2 regularization with $\lambda=10^{-6}$. Finally, we use Adam optimiser with a learning rate $\eta=10^{-4}$. Fig. \ref{fig:loss_epoch_curve} illustrates that the proposed MLP regressor learns the waveform translation task very efficiently, i.e., both the training loss and the validation loss steadily decrease as the number of epochs is increased.

\begin{figure}[ht]
\begin{center}
	\includegraphics[width=9cm, height=4.5cm]{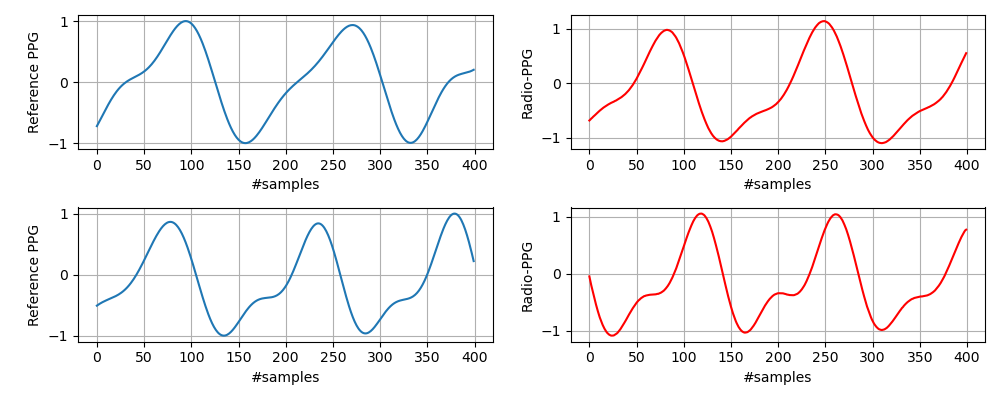} 
\caption{\footnotesize Two examples of synthetic PPG reconstruction from the radio data.}
\label{fig:ppg_synthetic}
\end{center}
\end{figure}

Moving forward, we first showcase the performance of the proposed method in a qualitative manner. Fig. \ref{fig:ppg_synthetic} illustrates by means of two graphical examples that the synthetic PPG reconstructed by the proposed method is very similar to reference PPG, albeit a few minor differences in morphology. That is, the synthetic and reference PPG signals are slightly different at the points of dicrotic notch. Further, there is a residual time offset between the two waveforms, but since this offset is really small (fraction of a second or less), it could be ignored by the clinicians. Nevertheless, it is worth mentioning that the synthetic PPG preserves a plurality of the clinically significant information, e.g., heart rate and more. This is what we believe makes the synthetic PPG waveform a valuable biomarker for clinicians and healthcare professionals for non-contact performance assessment of cardiovascular and respiratory systems of patients suffering from infectious diseases, e.g., covid19.

As for the quantitative results, we achieve a PPG waveform reconstruction MAE of 7.765 on the training set, and an MAE of 8.1294 on the test set. This is a pretty decent result, keeping in mind that the synthetic PPG waveform retains most of the morphological features of the reference PPG signal. 

%% file: conclusion.tex
\section{Conclusion \& Future Work}
\label{sec:conclusion}

This work proposed for the first time a novel non-contact method to monitor the PPG of a subject from a distance using radio signals in the microwave band. We utilized a pair of USRP SDRs whereby the transmit SDR exposed the chest of a subject with OFDM signals, while the receive SDR collected the modulated signal reflected off the chest of the subject. We collected 2.5 hours worth of raw radio data and reference PPG data from 16 subjects to train the proposed non-linear regression models. We conditioned the radio data and reference PPG data, and then passed it through the proposed pipeline in order to learn the PPG representation of the radio waveform. The proposed MLP regressor achieved a decent MAE of 8.1294. The synthetic PPG waveform has a great clinical significance as it could be used for non-contact performance assessment of cardiovascular and respiratory systems of patients suffering from infectious diseases, e.g., covid19.

As for the future work, we aim to study different fusion techniques (to optimally utilize the 64 OFDM sub-carriers) in order to improve the representation of the synthetic PPG signal. We believe this research opens new avenues for the development of contactless health sensing technologies. 

%We notice that the ridge regression approach works fine for local and constrained datasets because this framework require us to store the model variables per person, therefore we have one model per person. The model was learned through ridge regression formulation in matrix form. In terms of generalization this methodology becomes prohibitive and then we decided to move forward with modeling the regression using deep neural networks.
%The FFNN. Scale: it is easy to setup/schedule the algorithm to be trained with new samples from new subjects or new samples from the same person in order to take care of data drift. Generalization: with this approach we are able to build one model where, ideally, it will be a good balance between representation and generalization capabilities. Now, we have one model for which we claim that is useful for predict new instances of PPG.